\documentclass[aps,prl,article,groupedaddress,twocolumn]{revtex4-1}
\usepackage{graphicx}
\usepackage{amsmath}
\usepackage{xcolor}
\usepackage{siunitx}
\usepackage{ulem}

\draft 

\def\rc{$\alpha$-RuCl$_{3}$}
\def\rcK{K$_{0.5}$RuCl$_{3}$}

\begin{document}


\title{Nearest-neighbor Kitaev exchange blocked by charge order in electron doped \rc}

\author{A. Koitzsch,$^{1}$ C. Habenicht,$^{1}$ E. M\"uller,$^{1}$ M. Knupfer,$^{1}$ B. B\"uchner,$^{1}$ S. Kretschmer,$^{1, 2}$ \mbox{M. Richter},$^{1,3}$ J. van den Brink,$^{1,3}$ F. B\"orrnert,$^{1, 4}$ D. Nowak,$^{5}$ A. Isaeva,$^{5}$ and Th. Doert$^{5}$}

\affiliation
{$^{1}$IFW-Dresden, Helmholtzstr. 20, D-01069 Dresden, Germany\\
{$^{2}$Helmholtz Zentrum Dresden Rossendorf, Postfach 51 01 19, D-01314 Dresden, Germany\\}
{$^{3}$Dresden Center for Computational Materials Science (DCMS), D-01062 Dresden, Germany\\}
{$^{4}$Materialwissenschaftliche Elektronenmikroskopie,  Universit\"at Ulm, Albert-Einstein-Allee 11, D-89081 Ulm, Germany\\}
\mbox{$^{5}$Technische Universit\"at Dresden, Department of Chemistry and Food Chemistry, D-01062 Dresden, Germany\\}}

\date{\today}

\begin{abstract}
A quantum spin-liquid might be realized in \rc, a honeycomb-lattice magnetic material with substantial spin-orbit coupling. Moreover, \rc\ is a Mott insulator, which implies the possibility that novel exotic phases occur upon doping. Here, we study the electronic structure of this material when intercalated with potassium by photoemission spectroscopy, electron energy loss spectroscopy, and density functional theory calculations. We obtain a stable stoichiometry at K$_{0.5}$RuCl$_3$. This gives rise to a peculiar charge disproportionation into formally Ru$^{2+}$ (4$d^6$) and Ru$^{3+}$ (4$d^5$). Every Ru 4$d^5$ site with one hole in the $t_{2g}$ shell is surrounded by nearest neighbors of 4$d^6$ character, where the $t_{2g}$ level is full and magnetically inert. Thus, each type of Ru sites forms a triangular lattice and nearest-neighbor interactions of the original honeycomb are blocked.
\end{abstract}
\maketitle 

The quantum spin liquid (QSL) is an exotic state of matter which carries fractionalized excitations, completely different from the standard spin waves found for conventional magnetic order \cite{Balents2010}. The long search for a realization of this state has recently led to \rc, a layered honeycomb Mott-insulator with a 4$d^5$ configuration and substantial spin-orbit coupling \cite{Plumb2014, Banerjee2016}. From measurements of the magnetic excitation spectrum \cite{Banerjee2016, Sandilands2015b, Banerjee2017}, but also from other experiments \cite{Majumder2015, Sears2015, Johnson2015, Cao2016b, Sandilands2016, Koitzsch2016a}, and theory \cite{Jackeli2009, Kim2015a, Chaloupka2016, Yadav2016a, Nasu2016}, evidence is mounting that \rc\ is close to a Kitaev QSL, that is, a realization of the exactly solvable Kitaev model \cite{Kitaev2006}, with some modifications due to Heisenberg interactions \cite{Chaloupka2010, Chaloupka2013, Majumder2015, Kubota2015}. In particular, there are indications of a QSL state in an external magnetic field \cite{Yadav2016a, Baek2017a,Wolter2017a,Hentrich2017,Leahy2017a, Zheng2017}. This offers the opportunity to study the fractionalized excitations, most prominently Majorana Fermions and, possibly, to exploit the fact that they are protected from decoherence for quantum information processing schemes \cite{Kitaev2003}.   

The magnetic state is usually described within the framework of Heisenberg--Kitaev Hamiltonians. However, these attempts face the difficulty that the exchange parameters are not exactly known and higher order interactions (\textit{i.e.} beyond nearest neighbor) can be decisive \cite{Rousochatzakis2015, Yadav2016a, Winter2016, Sizyuk2016a}. This problem hinders a deeper understanding and 
theoretical progress of the field.

The QSL is the main driver of interest in \rc, but \rc\ is also a Mott-insulator. Doping a Mott-insulator often results in novel ground states with intriguing properties. Well known examples are the cuprates and manganates. Usually, also the magnetic order associated with the Mott-state reacts sensitively to doping. Therefore, doping \rc\ is a promising proposition for both: i) stabilizing new, interesting phases, and ii) probing the properties of the QSL.   

Here, we study the electronic structure of electron doped \rc. This is achieved by \textit{in situ} potassium intercalation. An apparent goal of such an approach is to reach a metallic phase. However, this has not been accomplished, similar to previous attempts by rubidium doping instead of potassium \cite{Zhou2016b}. Nevertheless, we show by a combination of electron energy loss spectroscopy (EELS), photoemission spectroscopy (PES), and density functional theory (DFT) that electron doping alters the ground state of \rc\ in a peculiar fashion. We observe a charge disproportionation which quenches the magnetic moment at every alternate Ru site and may serve as a platform to study the interplay of nearest neighbor and next-nearest neighbor Kitaev exchange. 

Platelet-like single crystals 
of \rc\ were grown by chemical vapor transport reactions. 
PES measurements were performed using a laboratory based system at room temperature after cleaving the crystals \textit{in situ}.
The EELS measurements in transmission have been conducted using thin films ($\textit{d} \approx$ 100 nm) at $T=20$ K.
Undoped crystals were intercalated with potassium \textit{in situ} by metal vapor from SAES dispensers.
The density functional (DF) calculations were performed with the all-electron full-potential local-orbital (FPLO) code \cite{Koepernik1999, FPLO}. 
See the Supplemental Material for further details.

\begin{figure}[t]
\includegraphics[width=0.95\linewidth]{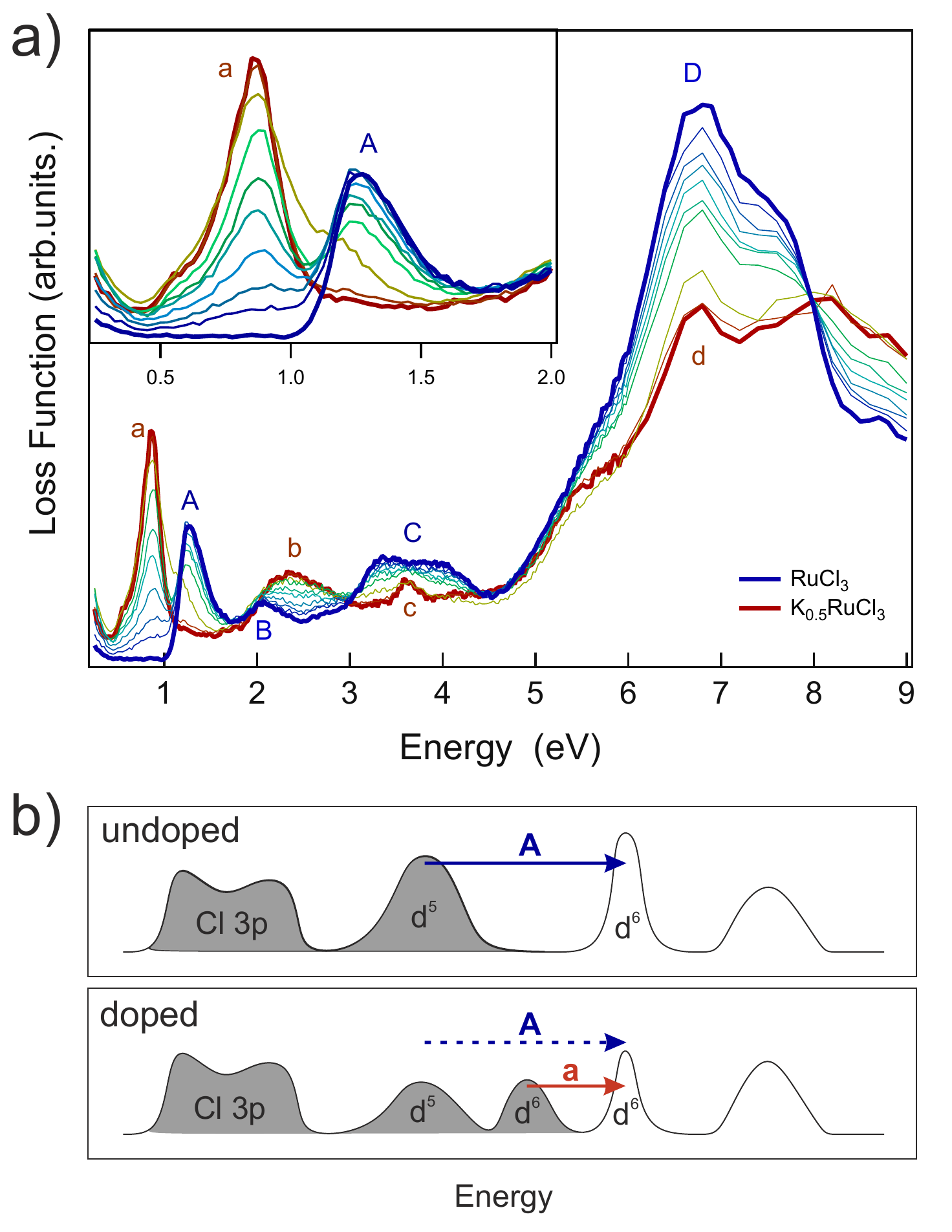}%
\caption{a) Loss function measured by electron energy loss spectroscopy for different degrees of potassium intercalation of \rc\ starting from the undoped material and reaching saturation at K$_{0.5}$RuCl$_3$. Capital letters mark peaks in the spectrum of the undoped, small letters of the doped material. (Inset) Magnification of the gap region. b) Schematic representation of the one-particle low-energy electronic structure of undoped and doped \rc. Capital and small letters correspond to the peak assignments in (a). 
}
\end{figure}

Figure 1a shows the effect of successive K intercalation on the low-energy loss function measured by EELS in transmission, a bulk sensitive probe. The pristine spectrum shows a peak at $E = 1.2$ eV (labeled \textbf{A}), which corresponds to optical gap excitations.
The peaks at higher energies (\textbf{B}--\textbf{D}) are due to crystal field and charge-transfer excitations \cite{Koitzsch2016a}. K intercalation causes drastic changes to the electronic structure, in particular, to the character of the gap. The inset of Fig. 1a expands the low-energy region. We observe that with increasing K content the spectral weight of the original \textbf{A} peak decreases and a new peak at lower energies appears (labeled \textbf{a}). Finally, at saturation, \textbf{A} completely vanishes. 
In order to find a rationale of these observations we show a schematic picture of the low-energy electronic structure in Fig. 1b. Near the Fermi energy the electronic states are dominated by Ru 4$d$ character. In the undoped case \textbf{A} corresponds to excitations across the Mott gap ($d^5d^5 \rightarrow d^4d^6$). With doping new $d^6$ states are created inside the gap. The fact that \textbf{A} completely vanishes in the experiment could be naturally explained by full electron doping, that is, every Ru$^{3+}$ (4$d^5$) ion is reduced to Ru$^{2+}$ (4$d^6$) by formation of KRuCl$_3$. Then the occupied $d^5$ states would disappear. However, this stoichiometry is not realized.

Quantitative X-ray photoemission spectroscopy (XPS), core level EELS, and DFT consistently hint at a saturated stoichiometry K$_{0.5}$RuCl$_3$ (see Supplemental Material).
Further, low-energy electron diffraction (LEED) shows a hexagonal pattern as in the pristine sample with modest lattice expansion but without superstructure (see Supplemental Material for details). 
This information allows us to construct a structural model where the K intercalation takes place between two adjacent Cl layers and occupies interstitial sites (see Fig. 2 and Supplemental Material for details). 
Most intriguingly, within our DF calculations a ground state with charge order among the two Ru sites,
denoted Ru(1) and Ru(2) henceforth, is found.
This charge order is not driven by structural symmetry breaking due to the K intercalation. To show this 
we relaxed the same structure, K$_{0.5}$RuCl$_3$(a), but with exchanged Ru charge and spin states, i.e.,
starting from the density and $d$-occupation data of the respective other Ru position (see SFig.1).
In this way we obtained a slightly modified geometry, denoted K$_{0.5}$RuCl$_3$(a') in
STab. I. The electronic state remained essentially unchanged, but with Ru(1)
and Ru(2) features exchanged.
The total energy of the new solution is only 4 meV per formula unit
higher than that of the previous one.
This means, the slightly different structural environment
of Ru(1) and Ru(2) in K$_{0.5}$RuCl$_3$(a) allows for charge and spin disproportionation due to the broken structural
symmetry; however, it does not determine which of the two Ru sites has a hole in the $t_{2g}$
band and which is non-magnetic.
Since the charge order even develops in the case of K$_{0.5}$RuCl$_3$(b) (see SFig.1), where
originally both Ru positions are identical, the order is generic for the considered composition
and structural details do not matter.

\begin{figure*}[t]
\includegraphics[width=0.9\linewidth]{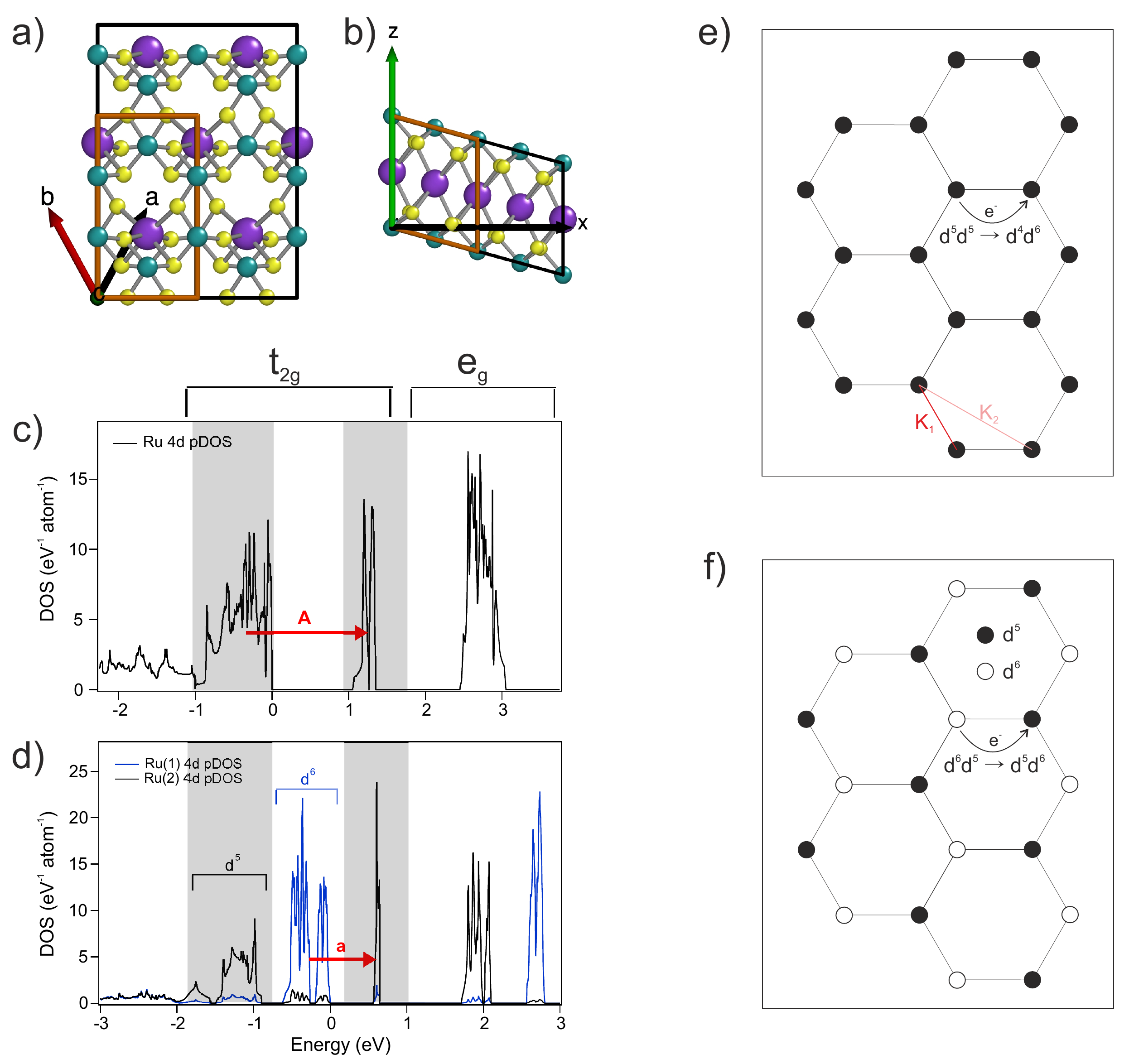}%
\caption{a, b) Proposed structure of K$_{0.5}$RuCl$_3$. The brown frame includes one conventional unit cell (four chemical units). The conventional b lattice vector is the sum of the indicated primitive lattice vectors a, b. 
(a) View onto the $a-b$ plane. 
(b) View onto the $x-z$ plane with distinct layers of Ru (green), Cl (yellow), and K (violet). Details of this and of another possible isomer, K$_{0.5}$RuCl$_3$(b), are given in Supplemental Material. 
c) Densities of states of pristine and d) K doped RuCl$_3$ with assignement of the gap feature of related EELS data. The DOS were obtained by means of full relativistics GGA+$U$ calculations in the ferro-magnetic state using relaxed structures detailled in the Supplemental Material. Here, both spin channels are added up. Grey shaded areas highlight lower and upper Hubbard band of the pristine RuCl$_3$ and how they are inherited to \rcK. e) Honeycomb lattice with $d^5$ configuration, corresponding to the undoped case. The Mott excitation is sketched. $K_1$ and $K_2$ denote nearest and next-nearest neighbor Kitaev exchange. f) Real space electronic structure of the fully doped K$_{0.5}$RuCl$_3$. 
}
\end{figure*}

The Ru $4d$ projected densities of states (DOS) of the isomer from Fig. 2a, b 
and of prisitine RuCl$_3$ are presented in Fig. 2c, d. 
The calculations support the schematic picture in Fig. 1b. New states are created inside the Mott gap by potassium doping.  
The striking difference between RuCl$_3$ and \rcK\ consists in a splitting of the occupied $t_{2g}$ and of the unoccupied $e_g$ bands in the latter case, caused by the charge order. The empty $t_{2g}$ sub-band of RuCl$_3$ can be identified as
driving force of the charge order: Upon doping this band to half-filling, it is split into an occupied, Ru(1) dominated and an unoccupied, Ru(2) dominated part. Further, the observed reduction of the gap size from \textbf{A} = 1.2 eV in RuCl$_3$
to \textbf{a} = 0.8 eV in \rcK\ can be understood by the splitting of the occupied $t_{2g}$ band
into an upper Ru(1) sub-band at the Fermi level and a lower Ru(2) sub-band. 
The distance of the latter to the first empty, Ru(2) band is fixed by the $U$ term.

The described situation is summarized in Fig. 2e, f and explains the complete suppression of \textbf{A} upon half doping.
In the undoped case, the gap is defined by the Mott excitation $d^5d^5\rightarrow d^4d^6$. 
In the doped case, this transition is blocked, as all $d^5$ sites are surrounded by $d^6$ sites.
The gap excitation \textbf{a} is now associated with the charge fluctuation $d^6d^5\rightarrow d^5d^6$.
Features \textbf{B}--\textbf{D} of the undoped material have been assigned recently to interband transitions with strong charge transfer character of \textbf{C} and \textbf{D} \cite{Koitzsch2016a} or, alternatively, to $d^4d^6$ multiplets (\textbf{A}--\textbf{C}) and charge transfer excitations (\textbf{D}) \cite{Sandilands2016b}.
While further work is needed for a unified theoretical description, we note that the $d^5$ initial state concentration still amounts to 50 \% under full doping and, thus, should be present in the doped spectrum but likewise reduced in intensity by 50 \%. This is indeed observed (see the charge transfer feature \textbf{d} in comparison to \textbf{D} in Fig. 1). 

Further quantitative support for the calculated charge-ordered ground state is provided by XPS core level spectra of the Ru 3$d$ states, Fig. 3.
The Ru 3$d$ line is spin-orbit split into $d_{5/2}$ and $d_{3/2}$ components. 
With doping a narrow lower-energy peak appears for both components. Quantitative analysis of these spectra is complicated 
and a complete fitting is not possible (see Supplemental Material).
However, we modeled the spectral shape of a high resolution scan of the fully doped sample (Fig. 3, main panel) by five Voigt peaks (two spin--orbit components for each of the two Ru sites plus one small charge transfer satellite at the high energy side) and a standard Shirley background. The intensity ratio of the spin--orbit components is fixed to $3:2$ and the Ru$^{2+}$ : Ru$^{3+}$ ratio to $1:1$. Peak positions and total width have been allowed to vary freely. This fit results in reasonable agreement with experiment, which implicitly confirms  a \rcK\ stoichiometry of the fully doped sample and the \mbox{Ru$^{2+}$ : Ru$^{3+}$} ratio.
The measured energy separation between Ru$^{2+}$ and Ru$^{3+}$ of $\Delta E = \SI{1.8}{eV}$ agrees well with the corresponding DF value of \SI{1.6}{eV}.

\begin{figure}[t]
\includegraphics[width=1\linewidth]{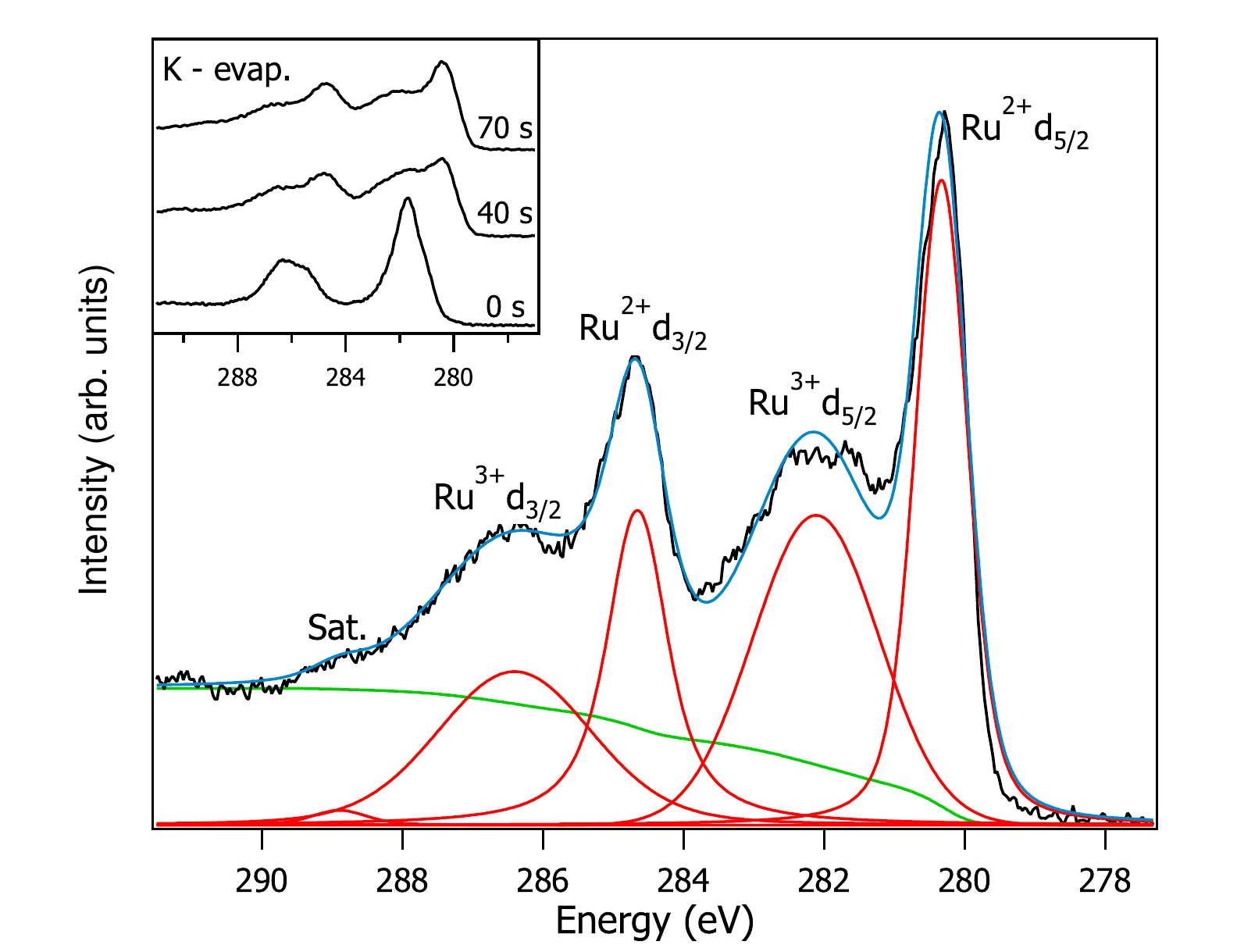}%
\caption{Ru 3$d$ core level measured by X-ray photoemission spectroscopy with model description. Inset: Dependence on K evaporation time.}%
\end{figure}

Nominally, K doping of \rc\ to \rcK\ reduces Ru(1) from Ru$^{3+}$ (4d$^5$) to Ru$^{2+}$ (4d$^6$), while Ru(2) remains in a (4d$^5$)
configuration. However, our Mulliken-type analysis tells a different story: the charge difference between Ru(1) and Ru(2) amounts to only 0.15 $e$ and the K electron is mainly supplied to the Cl ions being adjacent to the K layer. Almost 50 \% of the Ru $4d$ electrons form a broad hybrid band with Cl $3p$ states between $-$5.5 and $-$2.0 eV. Reversely, the narrow $t_{2g}$ and $e_g$ bands contain an appreachiable weight of Cl $3p$
between 10 \% and 60 \%, i.e., they are formed by $4d$-$3p$ molecular states of the appropriate symmetry (see Supplemental Material for details). 
Thus, a better description of the charge-ordered state would be K$^{1+}$[RuCl$_3$]$^0$[RuCl$_3$]$^{1-}$, where [RuCl$_3$]$^0$ has one hole in the $4d$-$3p$-$t_{2g}$ states and
[RuCl$_3$]$^{1-}$ has filled $4d$-$3p$-$t_{2g}$ states. Note, that each Cl ion is shared among one Ru(1) and one Ru(2); thus, all Cl ions carry almost the same charge.
For the sake of brevity, we will further stick to the nominal notation. 

\begin{figure*}
\includegraphics[width=0.9\linewidth]{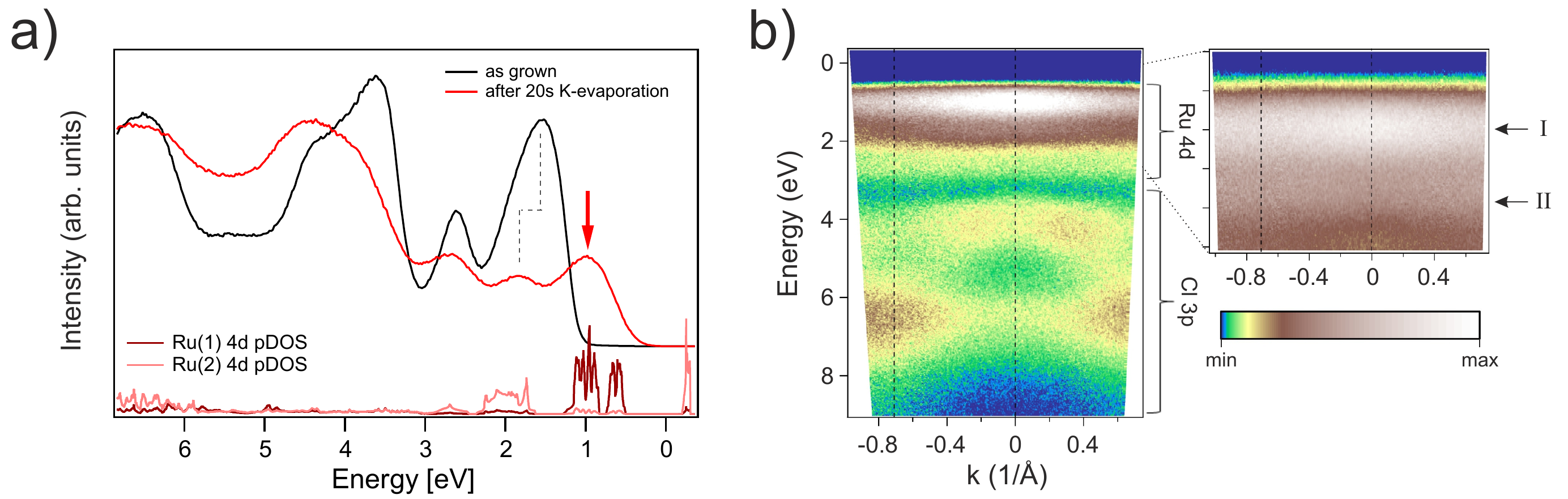}%
\caption{Valence band photoemission results. a) Comparison of undoped and electron doped \rc. The arrow marks the new, Ru$^{2+}$ ($d^6$) related peak and the dotted lines highlights the Ru$^{3+}$ states in the doped/undoped sample, respectively. Lower part: Ru 4$d$ partial density of states obtained by density functional theory. A stretching factor of 1.25 has been applied to the energy axis of the theory. b) angle dependence of the doped valence band and low-energy region.}
\end{figure*}

Figure 4 presents the valence band photoemission results.
The doped sample shows an additional peak at low energies (red arrow in Fig. 4a). 
The original Ru$^{3+}$ related peak is reduced in intensity and shifts somewhat to higher energies.
A similar spectral shape was obtained by Rb intercalation, although without any peak shift \cite{Zhou2016b}. 
A similar spectral evolution upon alkali metal doping of the $d^1$ Mott-insulator TiOCl has been explained previously by the electrostatic potential of the K$^+$ ions, which effectively localize the doped electron at the closest Ti site \cite{Sing2011}. 
In the present case, at variance, the charge order develops out of a rather
homogeneously distributed charge supplied by the K$^+$ ions to the Cl layers.

With increasing evaporation time the principal shape of the low-energy region does not change anymore (see the Supplemental Material). 
For better comparison of the DF calculations with experiment we have stretched the energy axis by a renormalization factor of 1.25 and shifted the energy by 0.5 eV within the gap. Such factors are often encountered when comparing photoemission data to DF calculations,
especially when the screening is bad \cite{Koitzsch2016a}.

Figure 4b shows the angle dependence of the valence band for the doped sample. The Cl 3$p$ related region shows clear dispersion, similar to the undoped sample \cite{Koitzsch2016a}. This excludes strong surface deterioration by the intercalation process. The Ru 4$d$ bands, on the other hand, do not show any dispersion, in agreement with previous results for Rb doping \cite{Zhou2016b}, although a small but finite dispersion is present for the undoped material \cite{Koitzsch2016a}. Within the charge disproportionation scheme this is readily understood: the Ru$^{3+}$ are too far apart in the doped sample to maintain a visible dispersion and so are the newly formed Ru$^{2+}$. The Cl network is homogeneously doped and retains its dispersion.

The charge ordered ground state found in the DF calculations
is consistent with EELS, valence band and core level photoemission data.
It is depcited in Fig. 2f. Other examples of charge order in half
doped Mott insulators are manganites, where the order is of checkerboard
type \cite{Tokura2000, Dagotto2001}.
Geometric frustration is present in magnetite (Fe$_3$O$_4$) and causes a complicated charge pattern below the Verwey transition \cite{Wright2002}. Transition metal dichalcogenides with similar crystal lattices tend to form charge density waves \cite{Rossnagel2011}. 
The observed real charge difference in charge ordered systems is always much smaller than one due to the large electrostatic energy cost.
Here, our DFT calculation yields a charge difference between Ru(1) and Ru(2) 
of only 0.15 $e$. Nevertheless, the local spin and orbital magnetic
moments are distinctly different:
$m_s=0.05\mu_B$, $m_l=0.11\mu_B$ for Ru(1), $m_s=0.73\mu_B$, $m_l=0.79\mu_B$ for Ru(2). This means, the charge order causes a difference of one order of magnitude between the magnetic moments of the two Ru sites.
The charge order develops due to the combination of two facts: (i) the number of electrons which are available for
the 4d-3p-$t_{2g}$ states of each \rc\ entity is not integer and (ii) a metallic state that would allow to distribute the extra charge equally among both Ru sites is prevented by the very small dispersion of the bands in the vicinity of the Fermi level. Reduction of $U$ decreases the gap until it is closed below $U = 1$ eV, $J = 0.2$ eV. Concomitantly, the charge order disappears and magnetic moments of  comparable size develop at the two Ru positions.

K intercalation offers a possibility to manipulate the charge and spin pattern of the honeycomb lattice in a controlled fashion. This is especially relevant because these patterns appear independently of the potassium lattice. The effect could be useful to study fundamental properties and to create qualitatively new magnetic groundstates. 
In particular, the zigzag antiferromagnetic order of the parent compound will disappear. The remaining effective triangular $d^5$ lattice for \rcK\ is geometrically frustrated. Model calculations are needed to elucidate the magnetic groundstate. As the dominating nearest-neighbor Kitaev exchange term vanishes the prominent background seen in INS \cite{Banerjee2016} and Raman \cite{Sandilands2015b} should decrease. Also the spin wave spectrum should vanish or, in case an alternative order is established, look different.
The suppresion of nearest neighbor Kitaev interactions might allow quantification of higher order terms, which is a prerequisite for a complete theoretical understanding of the QSL in \rc. In practice, the process of intercalation modifies bond lengths and, hence, the exchange parameters, which has to be taken into account for a quantitative description. But even on a qualitative level the originally homogeneous honeycomb lattice of \rc\ decomposes into two triangular lattices of $d^5$ and $d^6$ character which may host new magnetic groundstates \cite{Yadav2016a, Rousochatzakis2015}. 
Recently, in a different approach, the Ru honeycomb lattice has been diluted by Ir substitution, which quickly obstructs the antiferromagnetic order \cite{Lampen2016}.

In summary, we have investigated the electronic structure of potassium intercalated \rc. 
EELS, PES, and DFT show consistently and independently a stable \rcK\ stoichiometry in which a charge disproportionation into Ru$^{2+}$ and Ru$^{3+}$ takes place.
The charge order is accompanied by almost complete quenching of the magnetic moment at every alternate Ru site.
This type of combined charge and spin disproportionation on a honeycomb lattice is difficult to achieve otherwise. 
In principle, double perovskites may have some potential in this direction, but \rcK\ has the advantage of chemical simplicity.
The resulting peculiar state could offer a valuable platform for the investigation of the Kitaev exchange including higher-order interactions
and frustrated magnetism in general.

We thank U. Nitzsche for technical assistance and R. Ray and H. Rosner for valuable discussions. This work has been supported by the German Research Foundation (DFG) under SFB 1143.


\bibliography{D:/Literature/RuCl3}

\begin{thebibliography}{39}%
\makeatletter
\providecommand \@ifxundefined [1]{%
 \@ifx{#1\undefined}
}%
\providecommand \@ifnum [1]{%
 \ifnum #1\expandafter \@firstoftwo
 \else \expandafter \@secondoftwo
 \fi
}%
\providecommand \@ifx [1]{%
 \ifx #1\expandafter \@firstoftwo
 \else \expandafter \@secondoftwo
 \fi
}%
\providecommand \natexlab [1]{#1}%
\providecommand \enquote  [1]{``#1''}%
\providecommand \bibnamefont  [1]{#1}%
\providecommand \bibfnamefont [1]{#1}%
\providecommand \citenamefont [1]{#1}%
\providecommand \href@noop [0]{\@secondoftwo}%
\providecommand \href [0]{\begingroup \@sanitize@url \@href}%
\providecommand \@href[1]{\@@startlink{#1}\@@href}%
\providecommand \@@href[1]{\endgroup#1\@@endlink}%
\providecommand \@sanitize@url [0]{\catcode `\\12\catcode `\$12\catcode
  `\&12\catcode `\#12\catcode `\^12\catcode `\_12\catcode `\%12\relax}%
\providecommand \@@startlink[1]{}%
\providecommand \@@endlink[0]{}%
\providecommand \url  [0]{\begingroup\@sanitize@url \@url }%
\providecommand \@url [1]{\endgroup\@href {#1}{\urlprefix }}%
\providecommand \urlprefix  [0]{URL }%
\providecommand \Eprint [0]{\href }%
\providecommand \doibase [0]{http://dx.doi.org/}%
\providecommand \selectlanguage [0]{\@gobble}%
\providecommand \bibinfo  [0]{\@secondoftwo}%
\providecommand \bibfield  [0]{\@secondoftwo}%
\providecommand \translation [1]{[#1]}%
\providecommand \BibitemOpen [0]{}%
\providecommand \bibitemStop [0]{}%
\providecommand \bibitemNoStop [0]{.\EOS\space}%
\providecommand \EOS [0]{\spacefactor3000\relax}%
\providecommand \BibitemShut  [1]{\csname bibitem#1\endcsname}%
\let\auto@bib@innerbib\@empty
\bibitem [{\citenamefont {Balents}(2010)}]{Balents2010}%
  \BibitemOpen
  \bibfield  {author} {\bibinfo {author} {\bibfnamefont {L.}~\bibnamefont
  {Balents}},\ }\href {\doibase 10.1038/nature08917} {\bibfield  {journal}
  {\bibinfo  {journal} {Nature}\ }\textbf {\bibinfo {volume} {464}},\ \bibinfo
  {pages} {199} (\bibinfo {year} {2010})}\BibitemShut {NoStop}%
\bibitem [{\citenamefont {Plumb}\ \emph {et~al.}(2014)\citenamefont {Plumb},
  \citenamefont {Clancy}, \citenamefont {Sandilands}, \citenamefont {Shankar},
  \citenamefont {Hu}, \citenamefont {Burch}, \citenamefont {Kee},\ and\
  \citenamefont {Kim}}]{Plumb2014}%
  \BibitemOpen
  \bibfield  {author} {\bibinfo {author} {\bibfnamefont {K.~W.}\ \bibnamefont
  {Plumb}}, \bibinfo {author} {\bibfnamefont {J.~P.}\ \bibnamefont {Clancy}},
  \bibinfo {author} {\bibfnamefont {L.~J.}\ \bibnamefont {Sandilands}},
  \bibinfo {author} {\bibfnamefont {V.~V.}\ \bibnamefont {Shankar}}, \bibinfo
  {author} {\bibfnamefont {Y.~F.}\ \bibnamefont {Hu}}, \bibinfo {author}
  {\bibfnamefont {K.~S.}\ \bibnamefont {Burch}}, \bibinfo {author}
  {\bibfnamefont {H.-Y.}\ \bibnamefont {Kee}}, \ and\ \bibinfo {author}
  {\bibfnamefont {Y.-J.}\ \bibnamefont {Kim}},\ }\href {\doibase
  10.1103/PhysRevB.90.041112} {\bibfield  {journal} {\bibinfo  {journal} {Phys.
  Rev. B}\ }\textbf {\bibinfo {volume} {90}},\ \bibinfo {pages} {041112}
  (\bibinfo {year} {2014})}\BibitemShut {NoStop}%
\bibitem [{\citenamefont {Banerjee}\ \emph {et~al.}(2016)\citenamefont
  {Banerjee}, \citenamefont {Bridges}, \citenamefont {Yan}, \citenamefont
  {Aczel}, \citenamefont {Li}, \citenamefont {Stone}, \citenamefont {Granroth},
  \citenamefont {Lumsden}, \citenamefont {Yiu}, \citenamefont {Knolle},
  \citenamefont {Bhattacharjee}, \citenamefont {Kovrizhin}, \citenamefont
  {Moessner}, \citenamefont {Tennant}, \citenamefont {Mandrus},\ and\
  \citenamefont {Nagler}}]{Banerjee2016}%
  \BibitemOpen
  \bibfield  {author} {\bibinfo {author} {\bibfnamefont {A.}~\bibnamefont
  {Banerjee}}, \bibinfo {author} {\bibfnamefont {C.~A.}\ \bibnamefont
  {Bridges}}, \bibinfo {author} {\bibfnamefont {J.-Q.}\ \bibnamefont {Yan}},
  \bibinfo {author} {\bibfnamefont {A.~A.}\ \bibnamefont {Aczel}}, \bibinfo
  {author} {\bibfnamefont {L.}~\bibnamefont {Li}}, \bibinfo {author}
  {\bibfnamefont {M.~B.}\ \bibnamefont {Stone}}, \bibinfo {author}
  {\bibfnamefont {G.~E.}\ \bibnamefont {Granroth}}, \bibinfo {author}
  {\bibfnamefont {M.~D.}\ \bibnamefont {Lumsden}}, \bibinfo {author}
  {\bibfnamefont {Y.}~\bibnamefont {Yiu}}, \bibinfo {author} {\bibfnamefont
  {J.}~\bibnamefont {Knolle}}, \bibinfo {author} {\bibfnamefont
  {S.}~\bibnamefont {Bhattacharjee}}, \bibinfo {author} {\bibfnamefont {D.~L.}\
  \bibnamefont {Kovrizhin}}, \bibinfo {author} {\bibfnamefont {R.}~\bibnamefont
  {Moessner}}, \bibinfo {author} {\bibfnamefont {D.~A.}\ \bibnamefont
  {Tennant}}, \bibinfo {author} {\bibfnamefont {D.~G.}\ \bibnamefont
  {Mandrus}}, \ and\ \bibinfo {author} {\bibfnamefont {S.~E.}\ \bibnamefont
  {Nagler}},\ }\href {\doibase 10.1038/nmat4604} {\bibfield  {journal}
  {\bibinfo  {journal} {Nat Mater}\ }\textbf {\bibinfo {volume} {15}},\
  \bibinfo {pages} {733} (\bibinfo {year} {2016})}\BibitemShut {NoStop}%
\bibitem [{\citenamefont {Sandilands}\ \emph {et~al.}(2015)\citenamefont
  {Sandilands}, \citenamefont {Tian}, \citenamefont {Plumb}, \citenamefont
  {Kim},\ and\ \citenamefont {Burch}}]{Sandilands2015b}%
  \BibitemOpen
  \bibfield  {author} {\bibinfo {author} {\bibfnamefont {L.~J.}\ \bibnamefont
  {Sandilands}}, \bibinfo {author} {\bibfnamefont {Y.}~\bibnamefont {Tian}},
  \bibinfo {author} {\bibfnamefont {K.~W.}\ \bibnamefont {Plumb}}, \bibinfo
  {author} {\bibfnamefont {Y.-J.}\ \bibnamefont {Kim}}, \ and\ \bibinfo
  {author} {\bibfnamefont {K.~S.}\ \bibnamefont {Burch}},\ }\href {\doibase
  10.1103/PhysRevLett.114.147201} {\bibfield  {journal} {\bibinfo  {journal}
  {Phys. Rev. Lett.}\ }\textbf {\bibinfo {volume} {114}},\ \bibinfo {pages}
  {147201} (\bibinfo {year} {2015})}\BibitemShut {NoStop}%
\bibitem [{\citenamefont {Banerjee}\ \emph {et~al.}(2017)\citenamefont
  {Banerjee}, \citenamefont {Yan}, \citenamefont {Knolle}, \citenamefont
  {Bridges}, \citenamefont {Stone}, \citenamefont {Lumsden}, \citenamefont
  {Mandrus}, \citenamefont {Tennant}, \citenamefont {Moessner},\ and\
  \citenamefont {Nagler}}]{Banerjee2017}%
  \BibitemOpen
  \bibfield  {author} {\bibinfo {author} {\bibfnamefont {A.}~\bibnamefont
  {Banerjee}}, \bibinfo {author} {\bibfnamefont {J.}~\bibnamefont {Yan}},
  \bibinfo {author} {\bibfnamefont {J.}~\bibnamefont {Knolle}}, \bibinfo
  {author} {\bibfnamefont {C.~A.}\ \bibnamefont {Bridges}}, \bibinfo {author}
  {\bibfnamefont {M.~B.}\ \bibnamefont {Stone}}, \bibinfo {author}
  {\bibfnamefont {M.~D.}\ \bibnamefont {Lumsden}}, \bibinfo {author}
  {\bibfnamefont {D.~G.}\ \bibnamefont {Mandrus}}, \bibinfo {author}
  {\bibfnamefont {D.~A.}\ \bibnamefont {Tennant}}, \bibinfo {author}
  {\bibfnamefont {R.}~\bibnamefont {Moessner}}, \ and\ \bibinfo {author}
  {\bibfnamefont {S.~E.}\ \bibnamefont {Nagler}},\ }\href {\doibase
  10.1126/science.aah6015} {\bibfield  {journal} {\bibinfo  {journal}
  {Science}\ }\textbf {\bibinfo {volume} {356}},\ \bibinfo {pages} {1055}
  (\bibinfo {year} {2017})}\BibitemShut {NoStop}%
\bibitem [{\citenamefont {Majumder}\ \emph {et~al.}(2015)\citenamefont
  {Majumder}, \citenamefont {Schmidt}, \citenamefont {Rosner}, \citenamefont
  {Tsirlin}, \citenamefont {Yasuoka},\ and\ \citenamefont
  {Baenitz}}]{Majumder2015}%
  \BibitemOpen
  \bibfield  {author} {\bibinfo {author} {\bibfnamefont {M.}~\bibnamefont
  {Majumder}}, \bibinfo {author} {\bibfnamefont {M.}~\bibnamefont {Schmidt}},
  \bibinfo {author} {\bibfnamefont {H.}~\bibnamefont {Rosner}}, \bibinfo
  {author} {\bibfnamefont {A.~A.}\ \bibnamefont {Tsirlin}}, \bibinfo {author}
  {\bibfnamefont {H.}~\bibnamefont {Yasuoka}}, \ and\ \bibinfo {author}
  {\bibfnamefont {M.}~\bibnamefont {Baenitz}},\ }\href {\doibase
  10.1103/PhysRevB.91.180401} {\bibfield  {journal} {\bibinfo  {journal} {Phys.
  Rev. B}\ }\textbf {\bibinfo {volume} {91}},\ \bibinfo {pages} {180401}
  (\bibinfo {year} {2015})}\BibitemShut {NoStop}%
\bibitem [{\citenamefont {Sears}\ \emph {et~al.}(2015)\citenamefont {Sears},
  \citenamefont {Songvilay}, \citenamefont {Plumb}, \citenamefont {Clancy},
  \citenamefont {Qiu}, \citenamefont {Zhao}, \citenamefont {Parshall},\ and\
  \citenamefont {Kim}}]{Sears2015}%
  \BibitemOpen
  \bibfield  {author} {\bibinfo {author} {\bibfnamefont {J.~A.}\ \bibnamefont
  {Sears}}, \bibinfo {author} {\bibfnamefont {M.}~\bibnamefont {Songvilay}},
  \bibinfo {author} {\bibfnamefont {K.~W.}\ \bibnamefont {Plumb}}, \bibinfo
  {author} {\bibfnamefont {J.~P.}\ \bibnamefont {Clancy}}, \bibinfo {author}
  {\bibfnamefont {Y.}~\bibnamefont {Qiu}}, \bibinfo {author} {\bibfnamefont
  {Y.}~\bibnamefont {Zhao}}, \bibinfo {author} {\bibfnamefont {D.}~\bibnamefont
  {Parshall}}, \ and\ \bibinfo {author} {\bibfnamefont {Y.-J.}\ \bibnamefont
  {Kim}},\ }\href {\doibase 10.1103/PhysRevB.91.144420} {\bibfield  {journal}
  {\bibinfo  {journal} {Phys. Rev. B}\ }\textbf {\bibinfo {volume} {91}},\
  \bibinfo {pages} {144420} (\bibinfo {year} {2015})}\BibitemShut {NoStop}%
\bibitem [{\citenamefont {Johnson}\ \emph {et~al.}(2015)\citenamefont
  {Johnson}, \citenamefont {Williams}, \citenamefont {Haghighirad},
  \citenamefont {Singleton}, \citenamefont {Zapf}, \citenamefont {Manuel},
  \citenamefont {Mazin}, \citenamefont {Li}, \citenamefont {Jeschke},
  \citenamefont {Valent\'{\i}},\ and\ \citenamefont {Coldea}}]{Johnson2015}%
  \BibitemOpen
  \bibfield  {author} {\bibinfo {author} {\bibfnamefont {R.~D.}\ \bibnamefont
  {Johnson}}, \bibinfo {author} {\bibfnamefont {S.~C.}\ \bibnamefont
  {Williams}}, \bibinfo {author} {\bibfnamefont {A.~A.}\ \bibnamefont
  {Haghighirad}}, \bibinfo {author} {\bibfnamefont {J.}~\bibnamefont
  {Singleton}}, \bibinfo {author} {\bibfnamefont {V.}~\bibnamefont {Zapf}},
  \bibinfo {author} {\bibfnamefont {P.}~\bibnamefont {Manuel}}, \bibinfo
  {author} {\bibfnamefont {I.~I.}\ \bibnamefont {Mazin}}, \bibinfo {author}
  {\bibfnamefont {Y.}~\bibnamefont {Li}}, \bibinfo {author} {\bibfnamefont
  {H.~O.}\ \bibnamefont {Jeschke}}, \bibinfo {author} {\bibfnamefont
  {R.}~\bibnamefont {Valent\'{\i}}}, \ and\ \bibinfo {author} {\bibfnamefont
  {R.}~\bibnamefont {Coldea}},\ }\href {\doibase 10.1103/PhysRevB.92.235119}
  {\bibfield  {journal} {\bibinfo  {journal} {Phys. Rev. B}\ }\textbf {\bibinfo
  {volume} {92}},\ \bibinfo {pages} {235119} (\bibinfo {year}
  {2015})}\BibitemShut {NoStop}%
\bibitem [{\citenamefont {Cao}\ \emph {et~al.}(2016)\citenamefont {Cao},
  \citenamefont {Banerjee}, \citenamefont {Yan}, \citenamefont {Bridges},
  \citenamefont {Lumsden}, \citenamefont {Mandrus}, \citenamefont {Tennant},
  \citenamefont {Chakoumakos},\ and\ \citenamefont {Nagler}}]{Cao2016b}%
  \BibitemOpen
  \bibfield  {author} {\bibinfo {author} {\bibfnamefont {H.~B.}\ \bibnamefont
  {Cao}}, \bibinfo {author} {\bibfnamefont {A.}~\bibnamefont {Banerjee}},
  \bibinfo {author} {\bibfnamefont {J.-Q.}\ \bibnamefont {Yan}}, \bibinfo
  {author} {\bibfnamefont {C.~A.}\ \bibnamefont {Bridges}}, \bibinfo {author}
  {\bibfnamefont {M.~D.}\ \bibnamefont {Lumsden}}, \bibinfo {author}
  {\bibfnamefont {D.~G.}\ \bibnamefont {Mandrus}}, \bibinfo {author}
  {\bibfnamefont {D.~A.}\ \bibnamefont {Tennant}}, \bibinfo {author}
  {\bibfnamefont {B.~C.}\ \bibnamefont {Chakoumakos}}, \ and\ \bibinfo {author}
  {\bibfnamefont {S.~E.}\ \bibnamefont {Nagler}},\ }\href {\doibase
  10.1103/PhysRevB.93.134423} {\bibfield  {journal} {\bibinfo  {journal} {Phys.
  Rev. B}\ }\textbf {\bibinfo {volume} {93}},\ \bibinfo {pages} {134423}
  (\bibinfo {year} {2016})}\BibitemShut {NoStop}%
\bibitem [{\citenamefont {Sandilands}\ \emph
  {et~al.}(2016{\natexlab{a}})\citenamefont {Sandilands}, \citenamefont {Tian},
  \citenamefont {Reijnders}, \citenamefont {Kim}, \citenamefont {Plumb},
  \citenamefont {Kim}, \citenamefont {Kee},\ and\ \citenamefont
  {Burch}}]{Sandilands2016}%
  \BibitemOpen
  \bibfield  {author} {\bibinfo {author} {\bibfnamefont {L.~J.}\ \bibnamefont
  {Sandilands}}, \bibinfo {author} {\bibfnamefont {Y.}~\bibnamefont {Tian}},
  \bibinfo {author} {\bibfnamefont {A.~A.}\ \bibnamefont {Reijnders}}, \bibinfo
  {author} {\bibfnamefont {H.-S.}\ \bibnamefont {Kim}}, \bibinfo {author}
  {\bibfnamefont {K.~W.}\ \bibnamefont {Plumb}}, \bibinfo {author}
  {\bibfnamefont {Y.-J.}\ \bibnamefont {Kim}}, \bibinfo {author} {\bibfnamefont
  {H.-Y.}\ \bibnamefont {Kee}}, \ and\ \bibinfo {author} {\bibfnamefont
  {K.~S.}\ \bibnamefont {Burch}},\ }\href {\doibase 10.1103/PhysRevB.93.075144}
  {\bibfield  {journal} {\bibinfo  {journal} {Phys. Rev. B}\ }\textbf {\bibinfo
  {volume} {93}},\ \bibinfo {pages} {075144} (\bibinfo {year}
  {2016}{\natexlab{a}})}\BibitemShut {NoStop}%
\bibitem [{\citenamefont {Koitzsch}\ \emph {et~al.}(2016)\citenamefont
  {Koitzsch}, \citenamefont {Habenicht}, \citenamefont {M\"uller},
  \citenamefont {Knupfer}, \citenamefont {B\"uchner}, \citenamefont {Kandpal},
  \citenamefont {van~den Brink}, \citenamefont {Nowak}, \citenamefont
  {Isaeva},\ and\ \citenamefont {Doert}}]{Koitzsch2016a}%
  \BibitemOpen
  \bibfield  {author} {\bibinfo {author} {\bibfnamefont {A.}~\bibnamefont
  {Koitzsch}}, \bibinfo {author} {\bibfnamefont {C.}~\bibnamefont {Habenicht}},
  \bibinfo {author} {\bibfnamefont {E.}~\bibnamefont {M\"uller}}, \bibinfo
  {author} {\bibfnamefont {M.}~\bibnamefont {Knupfer}}, \bibinfo {author}
  {\bibfnamefont {B.}~\bibnamefont {B\"uchner}}, \bibinfo {author}
  {\bibfnamefont {H.~C.}\ \bibnamefont {Kandpal}}, \bibinfo {author}
  {\bibfnamefont {J.}~\bibnamefont {van~den Brink}}, \bibinfo {author}
  {\bibfnamefont {D.}~\bibnamefont {Nowak}}, \bibinfo {author} {\bibfnamefont
  {A.}~\bibnamefont {Isaeva}}, \ and\ \bibinfo {author} {\bibfnamefont
  {T.}~\bibnamefont {Doert}},\ }\href {\doibase 10.1103/PhysRevLett.117.126403}
  {\bibfield  {journal} {\bibinfo  {journal} {Phys. Rev. Lett.}\ }\textbf
  {\bibinfo {volume} {117}},\ \bibinfo {pages} {126403} (\bibinfo {year}
  {2016})}\BibitemShut {NoStop}%
\bibitem [{\citenamefont {Jackeli}\ and\ \citenamefont
  {Khaliullin}(2009)}]{Jackeli2009}%
  \BibitemOpen
  \bibfield  {author} {\bibinfo {author} {\bibfnamefont {G.}~\bibnamefont
  {Jackeli}}\ and\ \bibinfo {author} {\bibfnamefont {G.}~\bibnamefont
  {Khaliullin}},\ }\href {\doibase 10.1103/PhysRevLett.102.017205} {\bibfield
  {journal} {\bibinfo  {journal} {Phys. Rev. Lett.}\ }\textbf {\bibinfo
  {volume} {102}},\ \bibinfo {pages} {017205} (\bibinfo {year}
  {2009})}\BibitemShut {NoStop}%
\bibitem [{\citenamefont {Kim}\ \emph {et~al.}(2015)\citenamefont {Kim},
  \citenamefont {V.}, \citenamefont {Catuneanu},\ and\ \citenamefont
  {Kee}}]{Kim2015a}%
  \BibitemOpen
  \bibfield  {author} {\bibinfo {author} {\bibfnamefont {H.-S.}\ \bibnamefont
  {Kim}}, \bibinfo {author} {\bibfnamefont {V.~S.}\ \bibnamefont {V.}},
  \bibinfo {author} {\bibfnamefont {A.}~\bibnamefont {Catuneanu}}, \ and\
  \bibinfo {author} {\bibfnamefont {H.-Y.}\ \bibnamefont {Kee}},\ }\href
  {\doibase 10.1103/PhysRevB.91.241110} {\bibfield  {journal} {\bibinfo
  {journal} {Phys. Rev. B}\ }\textbf {\bibinfo {volume} {91}},\ \bibinfo
  {pages} {241110} (\bibinfo {year} {2015})}\BibitemShut {NoStop}%
\bibitem [{\citenamefont {Chaloupka}\ and\ \citenamefont
  {Khaliullin}(2016)}]{Chaloupka2016}%
  \BibitemOpen
  \bibfield  {author} {\bibinfo {author} {\bibfnamefont {J.}~\bibnamefont
  {Chaloupka}}\ and\ \bibinfo {author} {\bibfnamefont {G.}~\bibnamefont
  {Khaliullin}},\ }\href {\doibase 10.1103/PhysRevB.94.064435} {\bibfield
  {journal} {\bibinfo  {journal} {Phys. Rev. B}\ }\textbf {\bibinfo {volume}
  {94}},\ \bibinfo {pages} {064435} (\bibinfo {year} {2016})}\BibitemShut
  {NoStop}%
\bibitem [{\citenamefont {Yadav}\ \emph {et~al.}(2016)\citenamefont {Yadav},
  \citenamefont {Bogdanov}, \citenamefont {Katukuri}, \citenamefont
  {Nishimoto}, \citenamefont {van~den Brink},\ and\ \citenamefont
  {Hozoi}}]{Yadav2016a}%
  \BibitemOpen
  \bibfield  {author} {\bibinfo {author} {\bibfnamefont {R.}~\bibnamefont
  {Yadav}}, \bibinfo {author} {\bibfnamefont {N.~A.}\ \bibnamefont {Bogdanov}},
  \bibinfo {author} {\bibfnamefont {V.~M.}\ \bibnamefont {Katukuri}}, \bibinfo
  {author} {\bibfnamefont {S.}~\bibnamefont {Nishimoto}}, \bibinfo {author}
  {\bibfnamefont {J.}~\bibnamefont {van~den Brink}}, \ and\ \bibinfo {author}
  {\bibfnamefont {L.}~\bibnamefont {Hozoi}},\ }\href
  {http://dx.doi.org/10.1038/srep37925} {\bibfield  {journal} {\bibinfo
  {journal} {Scientific Reports}\ }\textbf {\bibinfo {volume} {6}},\ \bibinfo
  {pages} {37925 EP } (\bibinfo {year} {2016})}\BibitemShut {NoStop}%
\bibitem [{\citenamefont {Nasu}\ \emph {et~al.}(2016)\citenamefont {Nasu},
  \citenamefont {Knolle}, \citenamefont {Kovrizhin}, \citenamefont {Motome},\
  and\ \citenamefont {Moessner}}]{Nasu2016}%
  \BibitemOpen
  \bibfield  {author} {\bibinfo {author} {\bibfnamefont {J.}~\bibnamefont
  {Nasu}}, \bibinfo {author} {\bibfnamefont {J.}~\bibnamefont {Knolle}},
  \bibinfo {author} {\bibfnamefont {D.~L.}\ \bibnamefont {Kovrizhin}}, \bibinfo
  {author} {\bibfnamefont {Y.}~\bibnamefont {Motome}}, \ and\ \bibinfo {author}
  {\bibfnamefont {R.}~\bibnamefont {Moessner}},\ }\href
  {http://dx.doi.org/10.1038/nphys3809} {\bibfield  {journal} {\bibinfo
  {journal} {Nat Phys}\ }\textbf {\bibinfo {volume} {12}},\ \bibinfo {pages}
  {912} (\bibinfo {year} {2016})},\ \bibinfo {note} {letter}\BibitemShut
  {NoStop}%
\bibitem [{\citenamefont {Kitaev}(2006)}]{Kitaev2006}%
  \BibitemOpen
  \bibfield  {author} {\bibinfo {author} {\bibfnamefont {A.}~\bibnamefont
  {Kitaev}},\ }\href {\doibase http://dx.doi.org/10.1016/j.aop.2005.10.005}
  {\bibfield  {journal} {\bibinfo  {journal} {Annals of Physics}\ }\textbf
  {\bibinfo {volume} {321}},\ \bibinfo {pages} {2 } (\bibinfo {year}
  {2006})}\BibitemShut {NoStop}%
\bibitem [{\citenamefont {Chaloupka}\ \emph {et~al.}(2010)\citenamefont
  {Chaloupka}, \citenamefont {Jackeli},\ and\ \citenamefont
  {Khaliullin}}]{Chaloupka2010}%
  \BibitemOpen
  \bibfield  {author} {\bibinfo {author} {\bibfnamefont {J.}~\bibnamefont
  {Chaloupka}}, \bibinfo {author} {\bibfnamefont {G.}~\bibnamefont {Jackeli}},
  \ and\ \bibinfo {author} {\bibfnamefont {G.}~\bibnamefont {Khaliullin}},\
  }\href {\doibase 10.1103/PhysRevLett.105.027204} {\bibfield  {journal}
  {\bibinfo  {journal} {Phys. Rev. Lett.}\ }\textbf {\bibinfo {volume} {105}},\
  \bibinfo {pages} {027204} (\bibinfo {year} {2010})}\BibitemShut {NoStop}%
\bibitem [{\citenamefont {Chaloupka}\ \emph {et~al.}(2013)\citenamefont
  {Chaloupka}, \citenamefont {Jackeli},\ and\ \citenamefont
  {Khaliullin}}]{Chaloupka2013}%
  \BibitemOpen
  \bibfield  {author} {\bibinfo {author} {\bibfnamefont {J.}~\bibnamefont
  {Chaloupka}}, \bibinfo {author} {\bibfnamefont {G.}~\bibnamefont {Jackeli}},
  \ and\ \bibinfo {author} {\bibfnamefont {G.}~\bibnamefont {Khaliullin}},\
  }\href {\doibase 10.1103/PhysRevLett.110.097204} {\bibfield  {journal}
  {\bibinfo  {journal} {Phys. Rev. Lett.}\ }\textbf {\bibinfo {volume} {110}},\
  \bibinfo {pages} {097204} (\bibinfo {year} {2013})}\BibitemShut {NoStop}%
\bibitem [{\citenamefont {Kubota}\ \emph {et~al.}(2015)\citenamefont {Kubota},
  \citenamefont {Tanaka}, \citenamefont {Ono}, \citenamefont {Narumi},\ and\
  \citenamefont {Kindo}}]{Kubota2015}%
  \BibitemOpen
  \bibfield  {author} {\bibinfo {author} {\bibfnamefont {Y.}~\bibnamefont
  {Kubota}}, \bibinfo {author} {\bibfnamefont {H.}~\bibnamefont {Tanaka}},
  \bibinfo {author} {\bibfnamefont {T.}~\bibnamefont {Ono}}, \bibinfo {author}
  {\bibfnamefont {Y.}~\bibnamefont {Narumi}}, \ and\ \bibinfo {author}
  {\bibfnamefont {K.}~\bibnamefont {Kindo}},\ }\href {\doibase
  10.1103/PhysRevB.91.094422} {\bibfield  {journal} {\bibinfo  {journal} {Phys.
  Rev. B}\ }\textbf {\bibinfo {volume} {91}},\ \bibinfo {pages} {094422}
  (\bibinfo {year} {2015})}\BibitemShut {NoStop}%
\bibitem [{\citenamefont {Baek}\ \emph {et~al.}(2017)\citenamefont {Baek},
  \citenamefont {Do}, \citenamefont {Choi}, \citenamefont {Kwon}, \citenamefont
  {Wolter}, \citenamefont {Nishimoto}, \citenamefont {van~den Brink},\ and\
  \citenamefont {B\"uchner}}]{Baek2017a}%
  \BibitemOpen
  \bibfield  {author} {\bibinfo {author} {\bibfnamefont {S.-H.}\ \bibnamefont
  {Baek}}, \bibinfo {author} {\bibfnamefont {S.-H.}\ \bibnamefont {Do}},
  \bibinfo {author} {\bibfnamefont {K.-Y.}\ \bibnamefont {Choi}}, \bibinfo
  {author} {\bibfnamefont {Y.~S.}\ \bibnamefont {Kwon}}, \bibinfo {author}
  {\bibfnamefont {A.~U.~B.}\ \bibnamefont {Wolter}}, \bibinfo {author}
  {\bibfnamefont {S.}~\bibnamefont {Nishimoto}}, \bibinfo {author}
  {\bibfnamefont {J.}~\bibnamefont {van~den Brink}}, \ and\ \bibinfo {author}
  {\bibfnamefont {B.}~\bibnamefont {B\"uchner}},\ }\href {\doibase
  10.1103/PhysRevLett.119.037201} {\bibfield  {journal} {\bibinfo  {journal}
  {Phys. Rev. Lett.}\ }\textbf {\bibinfo {volume} {119}},\ \bibinfo {pages}
  {037201} (\bibinfo {year} {2017})}\BibitemShut {NoStop}%
\bibitem [{\citenamefont {Wolter}\ \emph {et~al.}(2017)\citenamefont {Wolter},
  \citenamefont {Corredor}, \citenamefont {Janssen}, \citenamefont {Nenkov},
  \citenamefont {Sch\"onecker}, \citenamefont {Do}, \citenamefont {Choi},
  \citenamefont {Albrecht}, \citenamefont {Hunger}, \citenamefont {Doert},
  \citenamefont {Vojta},\ and\ \citenamefont {B\"uchner}}]{Wolter2017a}%
  \BibitemOpen
  \bibfield  {author} {\bibinfo {author} {\bibfnamefont {A.~U.~B.}\
  \bibnamefont {Wolter}}, \bibinfo {author} {\bibfnamefont {L.~T.}\
  \bibnamefont {Corredor}}, \bibinfo {author} {\bibfnamefont {L.}~\bibnamefont
  {Janssen}}, \bibinfo {author} {\bibfnamefont {K.}~\bibnamefont {Nenkov}},
  \bibinfo {author} {\bibfnamefont {S.}~\bibnamefont {Sch\"onecker}}, \bibinfo
  {author} {\bibfnamefont {S.-H.}\ \bibnamefont {Do}}, \bibinfo {author}
  {\bibfnamefont {K.-Y.}\ \bibnamefont {Choi}}, \bibinfo {author}
  {\bibfnamefont {R.}~\bibnamefont {Albrecht}}, \bibinfo {author}
  {\bibfnamefont {J.}~\bibnamefont {Hunger}}, \bibinfo {author} {\bibfnamefont
  {T.}~\bibnamefont {Doert}}, \bibinfo {author} {\bibfnamefont
  {M.}~\bibnamefont {Vojta}}, \ and\ \bibinfo {author} {\bibfnamefont
  {B.}~\bibnamefont {B\"uchner}},\ }\href {\doibase 10.1103/PhysRevB.96.041405}
  {\bibfield  {journal} {\bibinfo  {journal} {Phys. Rev. B}\ }\textbf {\bibinfo
  {volume} {96}},\ \bibinfo {pages} {041405} (\bibinfo {year}
  {2017})}\BibitemShut {NoStop}%
\bibitem [{\citenamefont {{Hentrich}}\ \emph {et~al.}(2017)\citenamefont
  {{Hentrich}}, \citenamefont {{Wolter}}, \citenamefont {{Zotos}},
  \citenamefont {{Brenig}}, \citenamefont {{Nowak}}, \citenamefont {{Isaeva}},
  \citenamefont {{Doert}}, \citenamefont {{Banerjee}}, \citenamefont
  {{Lampen-Kelley}}, \citenamefont {{Mandrus}}, \citenamefont {{Nagler}},
  \citenamefont {{Sears}}, \citenamefont {{Kim}}, \citenamefont
  {{B{\"u}chner}},\ and\ \citenamefont {{Hess}}}]{Hentrich2017}%
  \BibitemOpen
  \bibfield  {author} {\bibinfo {author} {\bibfnamefont {R.}~\bibnamefont
  {{Hentrich}}}, \bibinfo {author} {\bibfnamefont {A.~U.~B.}\ \bibnamefont
  {{Wolter}}}, \bibinfo {author} {\bibfnamefont {X.}~\bibnamefont {{Zotos}}},
  \bibinfo {author} {\bibfnamefont {W.}~\bibnamefont {{Brenig}}}, \bibinfo
  {author} {\bibfnamefont {D.}~\bibnamefont {{Nowak}}}, \bibinfo {author}
  {\bibfnamefont {A.}~\bibnamefont {{Isaeva}}}, \bibinfo {author}
  {\bibfnamefont {T.}~\bibnamefont {{Doert}}}, \bibinfo {author} {\bibfnamefont
  {A.}~\bibnamefont {{Banerjee}}}, \bibinfo {author} {\bibfnamefont
  {P.}~\bibnamefont {{Lampen-Kelley}}}, \bibinfo {author} {\bibfnamefont
  {D.~G.}\ \bibnamefont {{Mandrus}}}, \bibinfo {author} {\bibfnamefont {S.~E.}\
  \bibnamefont {{Nagler}}}, \bibinfo {author} {\bibfnamefont {J.}~\bibnamefont
  {{Sears}}}, \bibinfo {author} {\bibfnamefont {Y.-J.}\ \bibnamefont {{Kim}}},
  \bibinfo {author} {\bibfnamefont {B.}~\bibnamefont {{B{\"u}chner}}}, \ and\
  \bibinfo {author} {\bibfnamefont {C.}~\bibnamefont {{Hess}}},\ }\href@noop {}
  {\bibfield  {journal} {\bibinfo  {journal} {ArXiv e-prints}\ } (\bibinfo
  {year} {2017})},\ \Eprint {http://arxiv.org/abs/1703.08623} {arXiv:1703.08623
  [cond-mat.str-el]} \BibitemShut {NoStop}%
\bibitem [{\citenamefont {Leahy}\ \emph {et~al.}(2017)\citenamefont {Leahy},
  \citenamefont {Pocs}, \citenamefont {Siegfried}, \citenamefont {Graf},
  \citenamefont {Do}, \citenamefont {Choi}, \citenamefont {Normand},\ and\
  \citenamefont {Lee}}]{Leahy2017a}%
  \BibitemOpen
  \bibfield  {author} {\bibinfo {author} {\bibfnamefont {I.~A.}\ \bibnamefont
  {Leahy}}, \bibinfo {author} {\bibfnamefont {C.~A.}\ \bibnamefont {Pocs}},
  \bibinfo {author} {\bibfnamefont {P.~E.}\ \bibnamefont {Siegfried}}, \bibinfo
  {author} {\bibfnamefont {D.}~\bibnamefont {Graf}}, \bibinfo {author}
  {\bibfnamefont {S.-H.}\ \bibnamefont {Do}}, \bibinfo {author} {\bibfnamefont
  {K.-Y.}\ \bibnamefont {Choi}}, \bibinfo {author} {\bibfnamefont
  {B.}~\bibnamefont {Normand}}, \ and\ \bibinfo {author} {\bibfnamefont
  {M.}~\bibnamefont {Lee}},\ }\href {\doibase 10.1103/PhysRevLett.118.187203}
  {\bibfield  {journal} {\bibinfo  {journal} {Phys. Rev. Lett.}\ }\textbf
  {\bibinfo {volume} {118}},\ \bibinfo {pages} {187203} (\bibinfo {year}
  {2017})}\BibitemShut {NoStop}%
\bibitem [{\citenamefont {{Zheng}}\ \emph {et~al.}(2017)\citenamefont
  {{Zheng}}, \citenamefont {{Ran}}, \citenamefont {{Li}}, \citenamefont
  {{Wang}}, \citenamefont {{Wang}}, \citenamefont {{Liu}}, \citenamefont
  {{Liu}}, \citenamefont {{Normand}}, \citenamefont {{Wen}},\ and\
  \citenamefont {{Yu}}}]{Zheng2017}%
  \BibitemOpen
  \bibfield  {author} {\bibinfo {author} {\bibfnamefont {J.}~\bibnamefont
  {{Zheng}}}, \bibinfo {author} {\bibfnamefont {K.}~\bibnamefont {{Ran}}},
  \bibinfo {author} {\bibfnamefont {T.}~\bibnamefont {{Li}}}, \bibinfo {author}
  {\bibfnamefont {J.}~\bibnamefont {{Wang}}}, \bibinfo {author} {\bibfnamefont
  {P.}~\bibnamefont {{Wang}}}, \bibinfo {author} {\bibfnamefont
  {B.}~\bibnamefont {{Liu}}}, \bibinfo {author} {\bibfnamefont
  {Z.}~\bibnamefont {{Liu}}}, \bibinfo {author} {\bibfnamefont
  {B.}~\bibnamefont {{Normand}}}, \bibinfo {author} {\bibfnamefont
  {J.}~\bibnamefont {{Wen}}}, \ and\ \bibinfo {author} {\bibfnamefont
  {W.}~\bibnamefont {{Yu}}},\ }\href@noop {} {\bibfield  {journal} {\bibinfo
  {journal} {ArXiv e-prints}\ } (\bibinfo {year} {2017})},\ \Eprint
  {http://arxiv.org/abs/1703.08474} {arXiv:1703.08474 [cond-mat.str-el]}
  \BibitemShut {NoStop}%
\bibitem [{\citenamefont {Kitaev}(2003)}]{Kitaev2003}%
  \BibitemOpen
  \bibfield  {author} {\bibinfo {author} {\bibfnamefont {A.}~\bibnamefont
  {Kitaev}},\ }\href {\doibase http://dx.doi.org/10.1016/S0003-4916(02)00018-0}
  {\bibfield  {journal} {\bibinfo  {journal} {Annals of Physics}\ }\textbf
  {\bibinfo {volume} {303}},\ \bibinfo {pages} {2 } (\bibinfo {year}
  {2003})}\BibitemShut {NoStop}%
\bibitem [{\citenamefont {Rousochatzakis}\ \emph {et~al.}(2015)\citenamefont
  {Rousochatzakis}, \citenamefont {Reuther}, \citenamefont {Thomale},
  \citenamefont {Rachel},\ and\ \citenamefont {Perkins}}]{Rousochatzakis2015}%
  \BibitemOpen
  \bibfield  {author} {\bibinfo {author} {\bibfnamefont {I.}~\bibnamefont
  {Rousochatzakis}}, \bibinfo {author} {\bibfnamefont {J.}~\bibnamefont
  {Reuther}}, \bibinfo {author} {\bibfnamefont {R.}~\bibnamefont {Thomale}},
  \bibinfo {author} {\bibfnamefont {S.}~\bibnamefont {Rachel}}, \ and\ \bibinfo
  {author} {\bibfnamefont {N.~B.}\ \bibnamefont {Perkins}},\ }\href {\doibase
  10.1103/PhysRevX.5.041035} {\bibfield  {journal} {\bibinfo  {journal} {Phys.
  Rev. X}\ }\textbf {\bibinfo {volume} {5}},\ \bibinfo {pages} {041035}
  (\bibinfo {year} {2015})}\BibitemShut {NoStop}%
\bibitem [{\citenamefont {Winter}\ \emph {et~al.}(2016)\citenamefont {Winter},
  \citenamefont {Li}, \citenamefont {Jeschke},\ and\ \citenamefont
  {Valent\'{\i}}}]{Winter2016}%
  \BibitemOpen
  \bibfield  {author} {\bibinfo {author} {\bibfnamefont {S.~M.}\ \bibnamefont
  {Winter}}, \bibinfo {author} {\bibfnamefont {Y.}~\bibnamefont {Li}}, \bibinfo
  {author} {\bibfnamefont {H.~O.}\ \bibnamefont {Jeschke}}, \ and\ \bibinfo
  {author} {\bibfnamefont {R.}~\bibnamefont {Valent\'{\i}}},\ }\href {\doibase
  10.1103/PhysRevB.93.214431} {\bibfield  {journal} {\bibinfo  {journal} {Phys.
  Rev. B}\ }\textbf {\bibinfo {volume} {93}},\ \bibinfo {pages} {214431}
  (\bibinfo {year} {2016})}\BibitemShut {NoStop}%
\bibitem [{\citenamefont {Sizyuk}\ \emph {et~al.}(2016)\citenamefont {Sizyuk},
  \citenamefont {W\"olfle},\ and\ \citenamefont {Perkins}}]{Sizyuk2016a}%
  \BibitemOpen
  \bibfield  {author} {\bibinfo {author} {\bibfnamefont {Y.}~\bibnamefont
  {Sizyuk}}, \bibinfo {author} {\bibfnamefont {P.}~\bibnamefont {W\"olfle}}, \
  and\ \bibinfo {author} {\bibfnamefont {N.~B.}\ \bibnamefont {Perkins}},\
  }\href {\doibase 10.1103/PhysRevB.94.085109} {\bibfield  {journal} {\bibinfo
  {journal} {Phys. Rev. B}\ }\textbf {\bibinfo {volume} {94}},\ \bibinfo
  {pages} {085109} (\bibinfo {year} {2016})}\BibitemShut {NoStop}%
\bibitem [{\citenamefont {Zhou}\ \emph {et~al.}(2016)\citenamefont {Zhou},
  \citenamefont {Li}, \citenamefont {Waugh}, \citenamefont {Parham},
  \citenamefont {Kim}, \citenamefont {Sears}, \citenamefont {Gomes},
  \citenamefont {Kee}, \citenamefont {Kim},\ and\ \citenamefont
  {Dessau}}]{Zhou2016b}%
  \BibitemOpen
  \bibfield  {author} {\bibinfo {author} {\bibfnamefont {X.}~\bibnamefont
  {Zhou}}, \bibinfo {author} {\bibfnamefont {H.}~\bibnamefont {Li}}, \bibinfo
  {author} {\bibfnamefont {J.~A.}\ \bibnamefont {Waugh}}, \bibinfo {author}
  {\bibfnamefont {S.}~\bibnamefont {Parham}}, \bibinfo {author} {\bibfnamefont
  {H.-S.}\ \bibnamefont {Kim}}, \bibinfo {author} {\bibfnamefont {J.~A.}\
  \bibnamefont {Sears}}, \bibinfo {author} {\bibfnamefont {A.}~\bibnamefont
  {Gomes}}, \bibinfo {author} {\bibfnamefont {H.-Y.}\ \bibnamefont {Kee}},
  \bibinfo {author} {\bibfnamefont {Y.-J.}\ \bibnamefont {Kim}}, \ and\
  \bibinfo {author} {\bibfnamefont {D.~S.}\ \bibnamefont {Dessau}},\ }\href
  {\doibase 10.1103/PhysRevB.94.161106} {\bibfield  {journal} {\bibinfo
  {journal} {Phys. Rev. B}\ }\textbf {\bibinfo {volume} {94}},\ \bibinfo
  {pages} {161106} (\bibinfo {year} {2016})}\BibitemShut {NoStop}%
\bibitem [{\citenamefont {Koepernik}\ and\ \citenamefont
  {Eschrig}(1999)}]{Koepernik1999}%
  \BibitemOpen
  \bibfield  {author} {\bibinfo {author} {\bibfnamefont {K.}~\bibnamefont
  {Koepernik}}\ and\ \bibinfo {author} {\bibfnamefont {H.}~\bibnamefont
  {Eschrig}},\ }\href {\doibase 10.1103/PhysRevB.59.1743} {\bibfield  {journal}
  {\bibinfo  {journal} {Phys. Rev. B}\ }\textbf {\bibinfo {volume} {59}},\
  \bibinfo {pages} {1743} (\bibinfo {year} {1999})}\BibitemShut {NoStop}%
\bibitem [{FPL()}]{FPLO}%
  \BibitemOpen
  \href@noop {} {}\bibinfo {howpublished}
  {\url{http://www.FPLO.de}}\BibitemShut {NoStop}%
\bibitem [{\citenamefont {Sandilands}\ \emph
  {et~al.}(2016{\natexlab{b}})\citenamefont {Sandilands}, \citenamefont {Sohn},
  \citenamefont {Park}, \citenamefont {Kim}, \citenamefont {Kim}, \citenamefont
  {Sears}, \citenamefont {Kim},\ and\ \citenamefont {Noh}}]{Sandilands2016b}%
  \BibitemOpen
  \bibfield  {author} {\bibinfo {author} {\bibfnamefont {L.~J.}\ \bibnamefont
  {Sandilands}}, \bibinfo {author} {\bibfnamefont {C.~H.}\ \bibnamefont
  {Sohn}}, \bibinfo {author} {\bibfnamefont {H.~J.}\ \bibnamefont {Park}},
  \bibinfo {author} {\bibfnamefont {S.~Y.}\ \bibnamefont {Kim}}, \bibinfo
  {author} {\bibfnamefont {K.~W.}\ \bibnamefont {Kim}}, \bibinfo {author}
  {\bibfnamefont {J.~A.}\ \bibnamefont {Sears}}, \bibinfo {author}
  {\bibfnamefont {Y.-J.}\ \bibnamefont {Kim}}, \ and\ \bibinfo {author}
  {\bibfnamefont {T.~W.}\ \bibnamefont {Noh}},\ }\href {\doibase
  10.1103/PhysRevB.94.195156} {\bibfield  {journal} {\bibinfo  {journal} {Phys.
  Rev. B}\ }\textbf {\bibinfo {volume} {94}},\ \bibinfo {pages} {195156}
  (\bibinfo {year} {2016}{\natexlab{b}})}\BibitemShut {NoStop}%
\bibitem [{\citenamefont {Sing}\ \emph {et~al.}(2011)\citenamefont {Sing},
  \citenamefont {Glawion}, \citenamefont {Schlachter}, \citenamefont {Scholz},
  \citenamefont {Go\ss{}}, \citenamefont {Heidler}, \citenamefont {Berner},\
  and\ \citenamefont {Claessen}}]{Sing2011}%
  \BibitemOpen
  \bibfield  {author} {\bibinfo {author} {\bibfnamefont {M.}~\bibnamefont
  {Sing}}, \bibinfo {author} {\bibfnamefont {S.}~\bibnamefont {Glawion}},
  \bibinfo {author} {\bibfnamefont {M.}~\bibnamefont {Schlachter}}, \bibinfo
  {author} {\bibfnamefont {M.~R.}\ \bibnamefont {Scholz}}, \bibinfo {author}
  {\bibfnamefont {K.}~\bibnamefont {Go\ss{}}}, \bibinfo {author} {\bibfnamefont
  {J.}~\bibnamefont {Heidler}}, \bibinfo {author} {\bibfnamefont
  {G.}~\bibnamefont {Berner}}, \ and\ \bibinfo {author} {\bibfnamefont
  {R.}~\bibnamefont {Claessen}},\ }\href {\doibase
  10.1103/PhysRevLett.106.056403} {\bibfield  {journal} {\bibinfo  {journal}
  {Phys. Rev. Lett.}\ }\textbf {\bibinfo {volume} {106}},\ \bibinfo {pages}
  {056403} (\bibinfo {year} {2011})}\BibitemShut {NoStop}%
\bibitem [{\citenamefont {Tokura}\ and\ \citenamefont
  {Nagaosa}(2000)}]{Tokura2000}%
  \BibitemOpen
  \bibfield  {author} {\bibinfo {author} {\bibfnamefont {Y.}~\bibnamefont
  {Tokura}}\ and\ \bibinfo {author} {\bibfnamefont {N.}~\bibnamefont
  {Nagaosa}},\ }\href {\doibase 10.1126/science.288.5465.462} {\bibfield
  {journal} {\bibinfo  {journal} {Science}\ }\textbf {\bibinfo {volume}
  {288}},\ \bibinfo {pages} {462} (\bibinfo {year} {2000})}\BibitemShut
  {NoStop}%
\bibitem [{\citenamefont {Dagotto}\ \emph {et~al.}(2001)\citenamefont
  {Dagotto}, \citenamefont {Hotta},\ and\ \citenamefont {Moreo}}]{Dagotto2001}%
  \BibitemOpen
  \bibfield  {author} {\bibinfo {author} {\bibfnamefont {E.}~\bibnamefont
  {Dagotto}}, \bibinfo {author} {\bibfnamefont {T.}~\bibnamefont {Hotta}}, \
  and\ \bibinfo {author} {\bibfnamefont {A.}~\bibnamefont {Moreo}},\ }\href
  {\doibase http://dx.doi.org/10.1016/S0370-1573(00)00121-6} {\bibfield
  {journal} {\bibinfo  {journal} {Physics Reports}\ }\textbf {\bibinfo {volume}
  {344}},\ \bibinfo {pages} {1 } (\bibinfo {year} {2001})}\BibitemShut
  {NoStop}%
\bibitem [{\citenamefont {Wright}\ \emph {et~al.}(2002)\citenamefont {Wright},
  \citenamefont {Attfield},\ and\ \citenamefont {Radaelli}}]{Wright2002}%
  \BibitemOpen
  \bibfield  {author} {\bibinfo {author} {\bibfnamefont {J.~P.}\ \bibnamefont
  {Wright}}, \bibinfo {author} {\bibfnamefont {J.~P.}\ \bibnamefont
  {Attfield}}, \ and\ \bibinfo {author} {\bibfnamefont {P.~G.}\ \bibnamefont
  {Radaelli}},\ }\href {\doibase 10.1103/PhysRevB.66.214422} {\bibfield
  {journal} {\bibinfo  {journal} {Phys. Rev. B}\ }\textbf {\bibinfo {volume}
  {66}},\ \bibinfo {pages} {214422} (\bibinfo {year} {2002})}\BibitemShut
  {NoStop}%
\bibitem [{\citenamefont {Rossnagel}(2011)}]{Rossnagel2011}%
  \BibitemOpen
  \bibfield  {author} {\bibinfo {author} {\bibfnamefont {K.}~\bibnamefont
  {Rossnagel}},\ }\href {http://stacks.iop.org/0953-8984/23/i=21/a=213001}
  {\bibfield  {journal} {\bibinfo  {journal} {Journal of Physics: Condensed
  Matter}\ }\textbf {\bibinfo {volume} {23}},\ \bibinfo {pages} {213001}
  (\bibinfo {year} {2011})}\BibitemShut {NoStop}%
\bibitem [{\citenamefont {{Lampen-Kelley}}\ \emph {et~al.}(2016)\citenamefont
  {{Lampen-Kelley}}, \citenamefont {{Banerjee}}, \citenamefont {{Aczel}},
  \citenamefont {{Cao}}, \citenamefont {{Yan}}, \citenamefont {{Nagler}},\ and\
  \citenamefont {{Mandrus}}}]{Lampen2016}%
  \BibitemOpen
  \bibfield  {author} {\bibinfo {author} {\bibfnamefont {P.}~\bibnamefont
  {{Lampen-Kelley}}}, \bibinfo {author} {\bibfnamefont {A.}~\bibnamefont
  {{Banerjee}}}, \bibinfo {author} {\bibfnamefont {A.~A.}\ \bibnamefont
  {{Aczel}}}, \bibinfo {author} {\bibfnamefont {H.~B.}\ \bibnamefont {{Cao}}},
  \bibinfo {author} {\bibfnamefont {J.-Q.}\ \bibnamefont {{Yan}}}, \bibinfo
  {author} {\bibfnamefont {S.~E.}\ \bibnamefont {{Nagler}}}, \ and\ \bibinfo
  {author} {\bibfnamefont {D.}~\bibnamefont {{Mandrus}}},\ }\href@noop {}
  {\bibfield  {journal} {\bibinfo  {journal} {ArXiv e-prints}\ } (\bibinfo
  {year} {2016})},\ \Eprint {http://arxiv.org/abs/1612.07202} {arXiv:1612.07202
  [cond-mat.str-el]} \BibitemShut {NoStop}%
\end{thebibliography}%

\end{document}